# A theory for calculating the number density distribution of small particles on a flat wall from pressure between the two walls

Kota Hashimoto[a] and Ken-ichi Amano[a]

[a]*Department of Ene*r*gy an*d *Hydrocarbon Chemistry, Graduate School of Engineering, K*yoto *University, Kyoto 615-8510, Japan.*

E-mail: hashimoto.kota.27z@st.kyoto-u.ac.jp

**ABSTRACT**

Surface force apparatus (SFA) and atomic force microscopy (AFM) can measure a force curve between a substrate and a probe in liquid. However, the force curve had not been transformed to the number density distribution of solvent molecules (colloidal particles) on the substance due to the absence of such a transform theory. Recently, we proposed and developed the transform theories for SFA and AFM. In these theories, the force curve is transformed to the pressure between two flat walls. Next, the pressure is transformed to number density distribution of solvent molecules (colloidal particles). However, pair potential between the solvent molecule (colloidal particle) and the wall is needed as the input of the calculation and Kirkwood superposition approximation is used in the previous theories. In this letter, we propose a new theory that does not use both the pair potential and the approximation. Instead, it makes use of a structure factor between solvent molecules (colloidal particles) which can be obtained by X-ray or neutron scattering.

**MAIN TEXT**

Structure of solid-liquid interface has relationship with physical properties and reactions at interfaces. Measurement of the density distribution of solvent molecules at the interface is important to understand them. Furthermore, the density distribution of colloid particles on solids or membranes is important for study of soft matter. Hereinafter, we call both colloid and solvent molecule "particle". To get information about density distribution of particles at the solid-liquid interface, surface force apparatus (SFA) [1] or atomic force microscope (AFM) [2] is used. SFA can measure forces between two crossed cylinders immersed in a liquid and AFM can measure forces between a substrate and a tip. This force contains solvation force, which is occurred by the overlap of liquid structures on the solid. SFA and AFM results are used to discuss the structure of solid-liquid interface, but their results are the forces between two surfaces, not the density distribution.

Our objective is to develop a method to transform the results obtained by SFA and AFM into the density distribution of particles. Recently K. Amano *et al*. [3] developed a method to transform the force from AFM into density distribution of particles and K. Hashimoto and K. Amano [4] developed a method to transform the force from SFA into density distribution of particles. In both papers the result from SFA/AFM is transformed to the pressure between two flat walls, and then the pressure is transformed to the density distribution of particles by using the Kirkwood superposition approximation and wall-particle pair potential [5, 6]. However, there are two problems in the method. The first problem is that the calculated density distribution has difference from the correct density distribution because of the usage of the Kirkwood superposition approximation. The Kirkwood superposition approximation has poor accuracy when the distance between the two walls are short and the bulk number density of the particles is high. This leads to discrepancy of density distributions at near the wall. The second problem is that the method needs the wall-particle pair potential which is hard to obtain. Approximation by hard wall potential can be used, but the wall-particle pair potential has an effect to the density distribution, which is not small. To overcome the problems, we propose a new method in this paper. In the new method, we use the hypernetted-chain (HNC) approximation [7] instead of the Kirkwood superposition approximation and the particle-particle radial distribution instead of wall-particle pair potential. HNC approximation is more accurate than the Kirkwood approximation [8]. Particle-particle radial distribution can be obtained by the structure factor from X-rays [9] or neutron scattering [10]. This new theory is related to a primitive theory proposed by K. Amano [11], however the primitive one can only calculate the number density distribution on a large spherical substrate. Furthermore, it utilizes a trial function of the number density distribution. Hence, the result is much dependent on the trial function. In the primitive version, the convergence method is also not sophisticated. On the other hand, the new theory can calculate without the trial function, and its convergence is efficient due to utilization of Newton-Raphson method.

The theory of the transform method is explained here. The model system contains two flat walls and infinite particles. The particle is considered to be a sphere. If the particle is a colloid particle, the solvent molecules are considered as continuum fluid, and their effects are included in the pair potentials. The two flat walls face each other and they are vertical to the $z$ axis. The shape of the walls are rectangular parallelepiped and the length of the sides are $X \times Y \times w$ which represents the length along $x \times y \times z$ axis. The area of the face which is vertical to the $z$ axis is expressed by $A$ (= $XY$). In the theory, the two walls are treated as big solutes, and the bulk number density of the walls is set as 0. The bulk number density of the particles is $\rho$.

Ornstein-Zernike equation of the system with two components (In this case, walls and particles) is expressed as [12],

$$\tilde{h}_{\text{WP}}(\mathbf{k}) - \tilde{c}_{\text{WP}}(\mathbf{k}) = \rho \tilde{c}_{\text{WP}}(\mathbf{k})\tilde{h}_{\text{PP}}(\mathbf{k}) = \tilde{h}_{\text{WP}}(\mathbf{k})\tilde{c}_{\text{PP}}(\mathbf{k}), \tag{1}$$

$$\tilde{\Gamma}_{\text{WW}}(\mathbf{k}) = \rho \tilde{h}_{\text{WP}}(\mathbf{k})\tilde{c}_{\text{WP}}(\mathbf{k}). \tag{2}$$

$h_{\text{WP}}$ is the wall-particle total correlation function (subscript WP stands for wall-particle), $c_{\text{WP}}$ is the wall-particle direct correlation function, $h_{\text{PP}}$ is the particle-particle total correlation function, $\Gamma_{\text{WW}}$ is the wall-wall indirect correlation function, $\mathbf{k}$ is an arbitrarily three dimensional vector in reverse space, $\tilde{f}$ is the Fourier transform of $f$. Substituting Eq. (1) into Eq. (2) yields,

$$\tilde{\Gamma}_{\text{WW}}(\mathbf{k}) = \frac{\rho \tilde{h}_{\text{WP}}(\mathbf{k})^2}{1 + \rho \tilde{h}_{\text{PP}}(\mathbf{k})}. \tag{3}$$

If $A$ is infinitely large, we can assume $\Gamma_{\text{WW}}(\mathbf{r})$ and $h_{\text{WP}}(\mathbf{r})$ depend on only $z$. $\mathbf{r}$ is an arbitrarily three dimensional vector in real space.

To calculate these functions in one dimension, we approximate that $\Gamma_{\text{WW}}$ and $h_{\text{WP}}$ depends on only $z$. This approximation is neglecting the effects of the side edges of the walls. To examine the property of these functions, we consider an arbitrary function $q$ which represents $\Gamma_{\text{WW}}$ or $h_{\text{WP}}$. $q$ is Fourier transformed in three dimensions,

$$\tilde{q}(\mathbf{k}) = \int_{-\frac{X}{2}}^{\frac{X}{2}} \exp(-ik_x r_x) \, dr_x \int_{-\frac{Y}{2}}^{\frac{Y}{2}} \exp(-ik_y r_y) \, dr_y \int_{-\infty}^{\infty} q(r_z) \exp(-ik_z r_z) \, dr_z. \tag{4}$$

The integrations with respect to $r_x$ and $r_y$ can be solved,

$$\tilde{q}(\mathbf{k}) = \frac{4 \sin\left(\frac{k_x X}{2}\right) \sin\left(\frac{k_y Y}{2}\right)}{k_x k_y} \int_{-\infty}^{\infty} q(r_z) \exp(-ik_z r_z) \, dr_z. \tag{5}$$

Taking the limit of $k_x \to 0$ and $k_y \to 0$ gives

$$\tilde{q}(0,0,k_z) = XY \int_{-\infty}^{\infty} q(r_z) \exp(-ik_z r_z)\, dr_z. \tag{6}$$

The integral part in Eq. (6) is the Fourier transform in one dimension. By using $XY = A$, we obtain

$$\tilde{q}(0,0,k_z) = A\tilde{q}(k_z). \tag{7}$$

Substituting $k_x = 0$ and $k_y = 0$ for Eq. (3) and using Eq. (7), we get

$$\frac{\tilde{\Gamma}_{\mathrm{WW}}(k_z)}{A} = \frac{\rho \tilde{h}_{\mathrm{WP}}(k_z)^2}{1 + \rho \tilde{h}_{\mathrm{PP}}(0,0,k_z)}. \tag{8}$$

$\tilde{h}_{\mathrm{PP}}$ is spherically symmetric and depends on only $k_r$. When $k_x = 0$ and $k_y = 0$, $k_r = k_z$, and then we obtain

$$\tilde{\Gamma}_{\mathrm{WW/A}}(k_z) = \frac{\rho \tilde{h}_{\mathrm{WP}}(k_z)^2}{1 + \rho \tilde{h}_{\mathrm{PP}}(k_z)}, \tag{9}$$

where we defined $\frac{\Gamma_{\mathrm{WW}}}{A} = \Gamma_{\mathrm{WW/A}}$. The force between the two walls $f$ (repulsion force is defined positive) has relation with wall-wall distribution function $g_{\mathrm{WW}}$ through wall-wall mean force potential $w$,

$$f(z) = -\frac{dw(z)}{dz} = \frac{d}{dz}\left(-\frac{1}{\beta} \ln g_{\mathrm{WW}}(z)\right), \tag{10}$$

where $\beta$ is inverse temperature $(= 1/k_{\mathrm{B}}T)$, $k_{\mathrm{B}}$ is Boltzmann constant, $T$ is absolute temperature. Hypernetted-chain approximation makes relation between $g_{\mathrm{WW}}(z)$ and $\Gamma_{\mathrm{WW}}(z)$,

$$g_{\mathrm{WW}}(z) = \exp\bigl(-\beta u_{\mathrm{WW}}(z) + \Gamma_{\mathrm{WW}}(z)\bigr), \tag{11}$$

where $u_{\mathrm{WW}}$ is the wall-wall pair potential. Substituting Eq. (11) into Eq. (10) and dividing with $A$, we get

$$\beta\left(u_{\mathrm{WW/A}}(z) - \int_{z}^{\infty} P(s)\, ds\right) = \Gamma_{\mathrm{WW/A}}(z), \tag{12}$$

where $P$ is the pressure between the two walls, and $\frac{u_{\mathrm{WW}}}{A} = u_{\mathrm{WW/A}}$. This equation is same as derived in [13], but the derivation is different. We notice that if we use PY approximation [7] instead of HNC approximation, a term which is not divided by area will appear. This means if we change the size of area, the relation between $P$ and $\Gamma_{\text{WW/A}}$ will change, and this is unphysical. Eq. (9), (12) relates $P$ (what we have) and $h_{\text{WP}}$ (what we want to know), but it is difficult to calculate $h_{\text{WP}}$ from $P$ for two reasons. First, we can't calculate $\tilde{\Gamma}_{\text{WW/A}}$ because we need $\Gamma_{\text{WW/A}}$ among all regions to perform Fourier transform on it. From Eq. (12) we can calculate $\Gamma_{\text{WW/A}}$ among the range where we know $P$, but we cannot know $\Gamma_{\text{WW/A}}$ where the two walls overlap. Second, it is difficult to decide the positive and negative of $\tilde{h}_{\text{WP}}$ from $\tilde{h}_{\text{WP}}^2$. To avoid these problems, we solve Eq. (9) in terms of $h_{\text{WP}}$ instead of $\tilde{h}_{\text{WP}}^2$. In this letter, we make use of Newton-Raphson method to obtain $h_{\text{WP}}$. To equalize the number of unknowns of $h_{\text{WP}}$ and knowns of $\Gamma_{\text{WW/A}}$, we set $h_{\text{WP}} = -1$ where the particle and the wall overlaps, and $h_{\text{WP}} = 0$ where the particle is far from the wall. (For example, if we know $\Gamma_{\text{WW/A}}$ at the range of $1\text{nm} \leq z \leq 10\text{nm}$, we will set $h_{\text{WP}} = -1$ at the range of $0.5\text{nm} \geq z$, and $h_{\text{WP}} = 0$ at the range of $10\text{nm} \leq z \leq 10.5\text{nm}$.) The Newton-Raphson method for this problem is expressed as

$$h_{\text{WP}}^{new}(z) = h_{\text{WP}}^{old}(z) - \left(\frac{\delta \Gamma_{\text{WW/A}}[h_{\text{WP}}^{old}](z')}{\delta h_{\text{WP}}^{old}(z)}\right)^{-1} \left(\Gamma_{\text{WW/A}}[h_{\text{WP}}^{old}](z') - \Gamma_{\text{WW/A}}^{\exp}(z')\right), \quad (13)$$

where $(\boldsymbol{B})^{-1}$ represents the inverse matrix of matrix $\boldsymbol{B}$, functional derivative of $(\delta C/\delta D)$ is the Jacobi matrix which is calculated from functional derivative of $C$ with respect to $D$, $\Gamma_{\text{WW/A}}^{\exp}$ is the $\Gamma_{\text{WW/A}}$ calculated from $P$, $\Gamma_{\text{WW/A}}[h_{\text{WP}}^{old}]$ is $\Gamma_{\text{WW/A}}$ calculated from Eq. (9) and $h_{\text{WP}}^{old}$. We can calculate $h_{\text{WP}}^{new}$ from $h_{\text{WP}}^{old}$ through Eq. (13). $h_{\text{WP}}^{new}$ is expected to be close to the correct $h_{\text{WP}}$ than $h_{\text{WP}}^{old}$. Functional derivative of $\Gamma_{\text{WW/A}}[h_{\text{WP}}^{old}]$ with respect to $h_{\text{WP}}^{old}$ is expressed as

$$\left(\frac{\delta \Gamma_{\text{WW/A}}[h_{\text{WP}}^{old}](z')}{\delta h_{\text{WP}}^{old}(z)}\right)$$
$$= \lim_{\epsilon \to 0} FT^{-1} \left(\frac{\rho\left(FT[h_{\text{WP}}^{old}(z') + \epsilon(\delta(z'-z) + \delta(z'+z))]\right)^2 - \rho \tilde{h}_{\text{WP}}^{old}(k_z)^2}{\epsilon\left(1 + \rho \tilde{h}_{\text{PP}}(k_z)\right)}\right), \quad (14)$$

where $FT$ and $FT^{-1}$ are the Fourier transform and its inverse version respectively. Since $h_{\text{WP}}^{old}$ is even function, two delta functions appear. Using Eq. (1) and the linearity of Fourier transform, we obtain

$$\left(\frac{\delta \Gamma_{\text{WW/A}}[h_{\text{WP}}^{old}](z')}{\delta h_{\text{WP}}^{old}(z)}\right) = FT^{-1}[2\rho \tilde{c}_{\text{WP}}(k_z) FT[\delta(z'-z) + \delta(z'+z)]], \quad (15)$$

where

$$\tilde{c}_{\text{WP}}(k_z) = \frac{\rho \tilde{h}_{\text{WP}}^{old}(k_z)}{1 + \rho \tilde{h}_{\text{PP}}(k_z)}. \quad (16)$$

Applying the convolution theorem, we obtain

$$\left(\frac{\delta \Gamma_{\text{WW/A}}[h_{\text{WP}}^{old}](z')}{\delta h_{\text{WP}}^{old}(z)}\right) = 2\rho\left(c_{\text{WP}}(z') * \left(\delta(z'-z) + \delta(z'+z)\right)\right) \quad (17)$$
$$= 2\rho\left(c_{\text{WP}}(z'-z) + c_{\text{WP}}(z'+z)\right),$$

where

$$c_{\text{WP}}(z') = FT^{-1}\left[\frac{\rho \tilde{h}_{\text{WP}}^{old}(k_z)}{1 + \rho \tilde{h}_{\text{PP}}(k_z)}\right]. \quad (18)$$

The Jacobi matrix can be calculated by Eq. (17). By using Eqs. (12), (13) and (17), we can calculate $h_{\text{WP}}$. A simple initial guess of $h_{\text{WP}}$ is 0 except in the overlap region where the value is -1.

To check the validity of the transform theory, we verified it in a computationally closed cycle. First, we calculated $\tilde{h}_{\text{PP}}$ between rigid particles by using both Ornstein-Zernike equation and hypernetted-chain approximation. Next, we calculated $h_{\text{WP}}$ on a rigid wall. This $h_{\text{WP}}$ is a benchmark function. The particle density distribution confined between two walls is calculated, and $P$ is calculated from it. Then, $h_{\text{WP}}$ is calculated from both $P$ and $\tilde{h}_{\text{PP}}$ through the Newton-Raphson method, and compared with the benchmark $h_{\text{WP}}$. Consequently, we found that the both $h_{\text{WP}}$ are very similar, which means the transform theory proposed here is valid. Similarly, we tested other wall-particle potentials: rigid wall with attractive part of Lennard-Jones potential and Lennard-Jones potential. In the former potential, the benchmark and calculated $h_{\text{WP}}$ were similar, but in the latter potential, the Newton-Raphson method did not converge. Now, we are working on to find why it does not converge when the short-range repulsion is not rigid.

In summary, we have proposed a transform theory for calculating the number density distribution of the particles on the wall from pressure between the two walls. We have written the derivation process of the transform theory in detail, and explained the result of the verification test of the transform theory. In the near future, we will show graphical results of the verification tests and put it into the practical use.

**ACKNOWLEDGEMENTS**

We appreciate Tetsuo Sakka (Kyoto University), Naoya Nishi (Kyoto University) for the useful advice, discussions, and data. This work was supported by "Grant-in-Aid for Young Scientists (B) from Japan Society for the Promotion of Science (15K21100)".